# The Crucial First Step in the Discovery of Millisecond Pulsars

A.C.S. Readhead
California Institute of Technology, Pasadena, CA 91125, USA
acr@caltech.edu

**ABSTRACT**

Millisecond pulsars, like pulsars, have led to major advances in many areas of astronomy and physics. The discovery in 2023 of a cosmological gravitational wave (GW) stochastic background using millisecond pulsar timing arrays has focused attention on the importance of millisecond pulsars, to both multi-messenger astronomy and cosmology, and for identifying the origin of the GW stochastic background, which is hypothesized to be due to supermassive black hole binaries (SMBHBs). Unlike pulsars, however, for which the details of the discovery are well-known, those of millisecond pulsars are not well known. In particular, the details of the first crucial step in the discovery of millisecond pulsars, namely the discovery of interplanetary scintillation (IPS) in the radio source 4C 21.53, are known only to the author. This article presents a first-hand account of this crucial first step, which resulted ultimately in the discovery of millisecond pulses from this object. A brief description of interplanetary and interstellar scintillation and scattering is given in the Appendix.

**INTRODUCTION**

The discovery of pulsars (Hewish, Bell, et al., 1968) had a profound effect on astronomy and physics, by opening up the study of very dense states of matter to observations for the first time. This discovery of pulsars, which, with remarkable prescience, were discussed in the discovery paper as possibly being due to neutron stars, made it appear far less outlandish and much more likely that black holes actually exist and might be observable. Amongst the many key findings to come out of pulsar studies, one of the most significant was the observation of the gravitational wave induced decrease in orbital period with time (Taylor et al., 1979) in the binary pulsar 1913+16, which won the Nobel Prize for Hulse and Taylor in 1993.

Given the enormous impact that the discovery of pulsars had already had on astrophysics and physics, it came as a complete surprise to the astronomical world when Backer, et al. (1982) discovered millisecond pulses from the radio source 4C 21.53, thereby opening up the new field of millisecond pulsars (Phinney and Kulkarni, 1994), which has in itself had a profound effect on astronomy over the last 4 decades, culminating just last year in the discovery of a stochastic background of cosmological gravitational waves (GW) (Agazie et al, 2023, EPTA Collaboration et al., 2023) which are thought to be produced by supermassive black hole binaries (SMBHBs).

With the rapidly growing importance of millisecond pulsars to astrophysics and cosmology, it is timely to give an account of the little-known first crucial step in the discovery of millisecond pulsars, without which the following steps would not have occurred, and for which purpose the author visited Lord's Bridge Observatory in April, 2024, in search of the interplanetary scintillation (IPS) records that first revealed the radio source 4C 21.53 as a scintillating source, thereby drawing the attention of the astronomy community to this object. In this article the history of the discovery of IPS in 4C 21.53 is presented for the first time.

It is illuminating of the misperceptions that so often arise in science that 731 papers (and counting) cite Backer, et al. (1982), whereas only 4 papers cite Readhead and Hewish (1974) in connection with millisecond pulsars. This, in spite of the fact that it was the IPS of 4C 21.53, which led Backer, et al. (1982), to observe it. Without this IPS result, the discovery of millisecond pulsars would have been delayed by some years, if not a decade or more. In an excellent, detailed, article on the discovery of millisecond pulsars, by Demorest et al. (2024) (JAHH, submitted), although the discovery of IPS reported by Readhead and Hewish (1974) is included, the critical pre-1976 period is not discussed in any detail. These two articles are therefore complementary, and the combination of the two provides a complete account of the discovery of millisecond pulsars.

It is illustrative of the difficulty of accurately tracing the history of discoveries, to note that a critical misperception - that the 4C survey (Pilkington and Scott, 1965; Gower, et al., 1967) was *required* for Readhead and Hewish (1974) to identify and draw attention to the compact source that Backer, et al. (1982) observed - is even propagated, unwittingly, in the article by Demorest et al. (2024). In fact the 4C catalog was not required at all, as explained in the Conclusion of this article. This shows just how difficult it is to get the history right. At this point the author is switching to the first person for easier flow of the narrative.

During the period from 1976 to 1982 Don Backer and I had many meetings, and many discussions, because we were both deeply involved in the development of the US Very Long Baseline Interferometry (VLBI) Network, which was being led by Marshall Cohen. As young post-docs, a lot of the burden of scheduling the VLBI observations, setting up the telescopes, making the observations, correlating the data, and analyzing the results, fell on our shoulders, and on occasion we observed together at the Hat Creek Observatory or the Owens Valley Radio Observatory, although we were more often both stationed at our own observatories, and in frequent telephone contact as we set up the VLBI observations. We did not write many papers together because we had our own scientific projects, but we worked on most US VLBI network sessions together. During this period we had frequent discussions about the nature of 4C 21.53, and we both were convinced that the most likely explanation was that it is a peculiar pulsar. We had both decided to follow it up at Arecibo, and I was preparing a proposal for time at Arecibo, with Caltech graduate student Rick Edelson, when I heard of the discovery of millisecond pulses from 4C 21.53 by Backer, et al. (1982).

**INTERPLANETARY SCINTILLATION OBSERVATIONS AT THE CAVENDISH FROM 1967 TO 1976**

Detailed accounts of the IPS (Hewish, et al. 1964) program at the Cavendish Laboratory from 1967 to 1976 may be found in the PhD theses of Leslie Little (Little, 1968), Jocelyn Bell-Burnell (Bell, 1968}, myself (Readhead, 1972}, and Peter Duffett-Smith (Duffett-Smith, (1976). This account focuses on my own thesis.

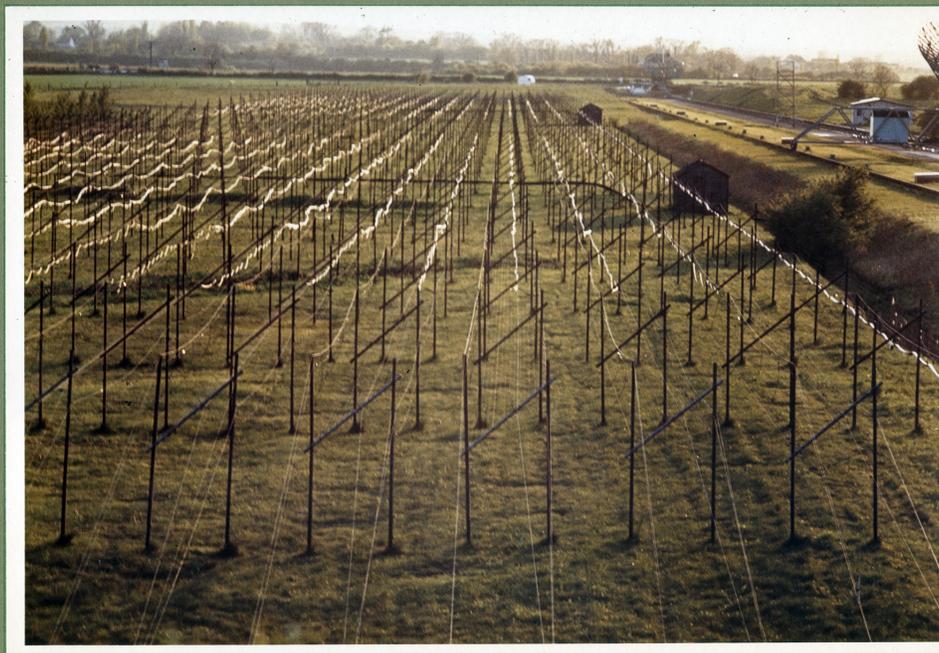

*Figure 1: Photo of the 4-acre Telescope from the author's thesis (Readhead, 1972). The east-west dimension >> the north-south dimension, so the beam is much wider in declination than in right ascension. There were no drones in those days, so the author had to climb a tree to get this perspective of the antenna. The author chose late afternoon, when the low sun reflecting off the cables made the structure more visible.*

When I arrived at the Cavendish Laboratory in early September 1968, the first person I met there was Jocelyn Bell, who was packing up preparatory to leaving Cambridge.  Being midsummer, there were few people around in the Mullard Radio Astronomy Group, Martin Ryle's group at the Cavendish Laboratory, so Bell very kindly took me under her wing and undertook to introduce me around and help me get settled.  She gave me her desk, and bequeathed to me her chart records, and introduced me to Antony Hewish, her adviser.  I was particularly interested to meet Bell and Hewish because their paper on the discovery of pulsars (Hewish, Bell, et al., 1968) had led me, in February of that year, to switch, from applying to optical astronomy, to applying to radio astronomy, at Cambridge. My main interest was quasars, a topic on which, in 1964, I had given my first scientific talk as a physics undergraduate at the University of the Witwatersrand, and the best way of studying large numbers of quasars was by using the interplanetary scintillation technique (Hewish, et al. 1964).

I asked Bell whether she had any recommendations on the choice of advisers and she strongly advised me to work with Hewish, because, as she put it, "You can't do better than Hewish".  However, the Cambridge One Mile Telescope had just come into operation, and I was fascinated by aperture synthesis, and so requested to work on quasars with Martin Ryle, using the One Mile Telescope. I knew that another incoming research student, the term used for graduate students in the UK, namely Simon Mitton, had also requested to work with Ryle on quasars.  At the faculty meeting to discuss the allocation of research projects amongst the new students Ryle chose Mitton, and Hewish informed me with the words "I'm afraid you've got me".  Both Bell and Hewish thought it would be a good idea for me to go through Bell's PhD survey IPS charts to see whether she had missed any pulsars, which I did.  It took several weeks for me to determine that Bell had not missed any pulsars. Bell and I remained in close touch following her departure from Cambridge because she and Hewish were writing a paper (Hewish and Burnell, 1970) on the observations, and I was working on the theory of IPS, deriving the relationship between scintillation index and solar elongation (Readhead, 1971) to determine the angular sizes of the scintillating objects, which was used by them in their paper.

Hewish suggested that I should carry out a survey with the 4-acre Telescope (see Fig. 1) for IPS over the whole sky north of declination -8°, since Bell's IPS survey had covered only Right Ascensions $10^h$-$16^h$, and Declinations -8° to 44°. I at first accepted that this would be my thesis topic, but after a week thinking about it, I felt that IPS was a very roundabout way of studying quasars, because it required a deep dive into the theory of diffraction in irregular media and the properties of the solar wind, and I told Hewish that I was disenchanted with my proposed thesis.  Hewish responded "Are you telling me that you can't think of something interesting to do with this brand new telescope?". One of the most stimulating things about working with both Ryle and Hewish, was that the ball was so often back in your court before you had recovered from your service. The challenge was exactly what I needed.  Over a year later, Hewish and I were at our regular morning coffee break in the Austin wing of the Cavendish, when Hewish started quizzing me about the value and significance of my work, and asking me a lot of deep questions about it.  We spent a good 20 minutes talking about this, and, as we walked back to our offices, I became uneasy about all the questions, and said "Tony – all these questions – are you concerned that this project may not be worth doing?".  Hewish responded "Me? You're the one who was disenchanted. I've always known it is worth doing, I just wanted to be sure that you did too".  I was struck by my advisor's wisdom - knowing that only by becoming deeply involved in research could I become convinced of its value - and his patience - allowing me a year to gain that experience before bringing the subject up again.

Both Bell and Hewish strongly recommended that I should build a new "scintillation meter".  This  was needed because, in the original scintillation meter on the 4-acre Telescope, the gain on the critical high pass filtered, rectified, and  integrated output was too low for the noise  to be visible.  In addition, as both Bell and Hewish informed me, the original scintillation meter was ``full of 'dry joints' '', i.e. bad solder connections. I therefore dutifully set about building a new scintillation meter as my first real task as a research student.   In this I copied the overall design of the original scintillation meter, but I designed an amplifier with significantly higher gain than the original amplifier for the all-important rectified and integrated trace. The 4-acre telescope was equipped to observe four declination beams simultaneously, so I built four scintillation meters.  Early in 1969, a transmitting service, operating at 85.7 MHz was set up by a local company and it was necessary to suppress the resulting interference. I therefore designed and built filters with a 1.5 MHz bandpass, centered on 81.5 MHz with an attenuation of > 30 db at 85.7 MHz, which successfully eliminated the interference.

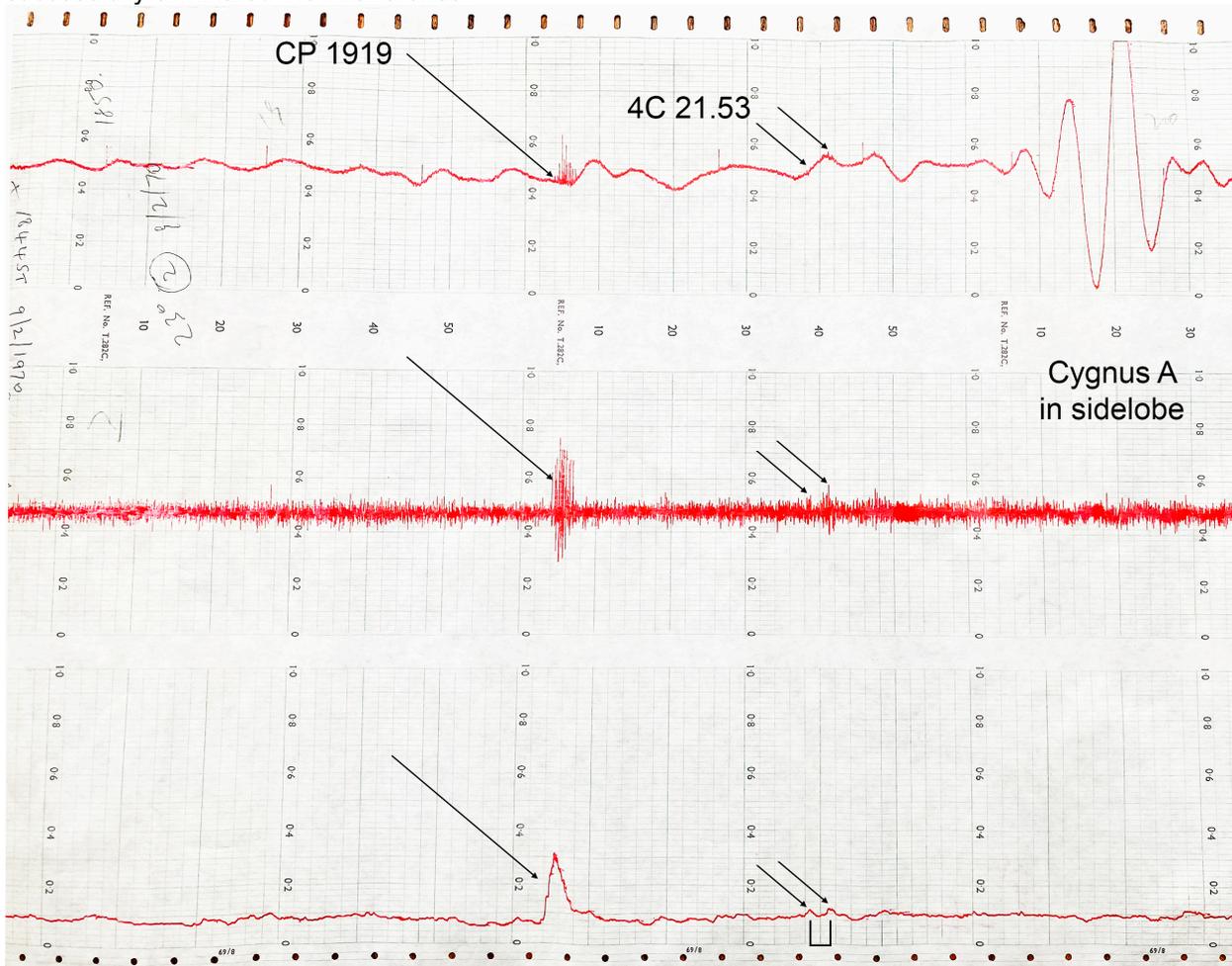

*Figure 2. A typical chart recording produced by the author's scintillation meter. The vertical axis is the flux density and the horizontal axis is the time, with the pips on the top trace indicating 10 minute intervals. The top trace shows the raw signal from the phase-switching receiver; the middle trace shows the high-pass filtered record; the bottom trace shows the rectified and integrated record, in which the vertical axis is the scintillating flux density. This is the best surviving chart of IPS in 4C 21.53. It was observed on 9 February, 1970, when the solar elongation of 4C 21.53 was optimal (45.3º). Two arrows mark 4C 21.53 because the source exhibited two peaks as it transited the beam of the telescope, the centroids of the lobes are indicated by the black U shape, and show that the separation of the peaks is exactly as expected. As explained in the text, this is a key signature of objects exhibiting IPS on the 4-acre Telescope. Transit through the third (weaker) lobe was not seen. The first pulsar, CP 1919, is also seen here, showing strong pulses of emission.*

I installed the new scintillation meters at Lord's Bridge Observatory on 10 May, 1969 (see Fig. 3). The 4-acre Telescope observing log is preserved in the Hewish archives at Churchill College in Cambridge, and the first entry in my hand, on 10 May 1969, has the heading "What went wrong when, and how we tried to fix it". The rest of the log covers the period of my IPS survey (Readhead, 1972) each time I carried out tests on the 4-acre Telescope and receiver system. A typical record of the output of one of my scintillation meters is shown in Fig. 2. There are three traces in red ink on the green chart paper. The writing at the extreme left is the technician's (Reg Dye or Don Rolph) note of when he started the new chart running. The other writing is in my hand. The date is 9 February 1970. This is the best surviving chart record of IPS in 4C 21.53 from my IPS survey, preserved thanks to John Baldwin, who chose to preserve some charts. The first pulsar, CP 1919, is clearly visible (Hewish, Bell, et al., 1968).

The top trace shows the "raw record", which is the output from the 4-acre Telescope's 81.5 MHz phase-switching receiver system. The amplitude of the signal is proportional to the flux density of the source. In the scintillation meter, time pips every 10 minutes are added to this trace, and a 50 Hz "anti-stick" signal is added to all three traces to ensure that the pens do not get stuck on the chart paper. The middle trace shows the high-pass filtered and amplified record, and the bottom trace shows the rectified and integrated high-pass filtered record. My IPS survey produced 5 miles of these chart records. This chart record from 9 February 1970 is the only surviving record we have in which 4C 21.53 can be seen (just) by the practiced eye in all three traces. At this time the solar elongation of 4C 21.53 is at its minimum (45.3°), which is when the IPS is strongest. So this is at the level of the strongest IPS seen in 4C 21.53 in my survey (Readhead, 1972). In view of how weak the scintillating signal was, and because of the misidentification of some non-scintillating sources as scintillators discussed later in this article, at this time I was not convinced that this was a true scintillating signal, but the two peaks seen in 4C 21.53 have the same separation in time as that between the sidelobe and the main beam, as can be seen from the black U-shape below the bottom trace, which shows when 4C 21.53 transited the first sidelobe and then the main beam. The following sidelobe was weaker and did not register in the scintillation meter. A small adjustment in the cable lengths from the two halves of the antenna to the phase-switching receiver would be needed to correct this. On the other surviving records, 4C 21.53 is visible only on the bottom trace, i.e. the rectified and integrated high-pass filtered record, and was totally invisible on the other two traces. Since it is comparable to the noise, the signal on the integrated (bottom) trace would therefore have been invisible with the original scintillation meter, rendering the IPS in 4C 21.53 almost impossible to see.

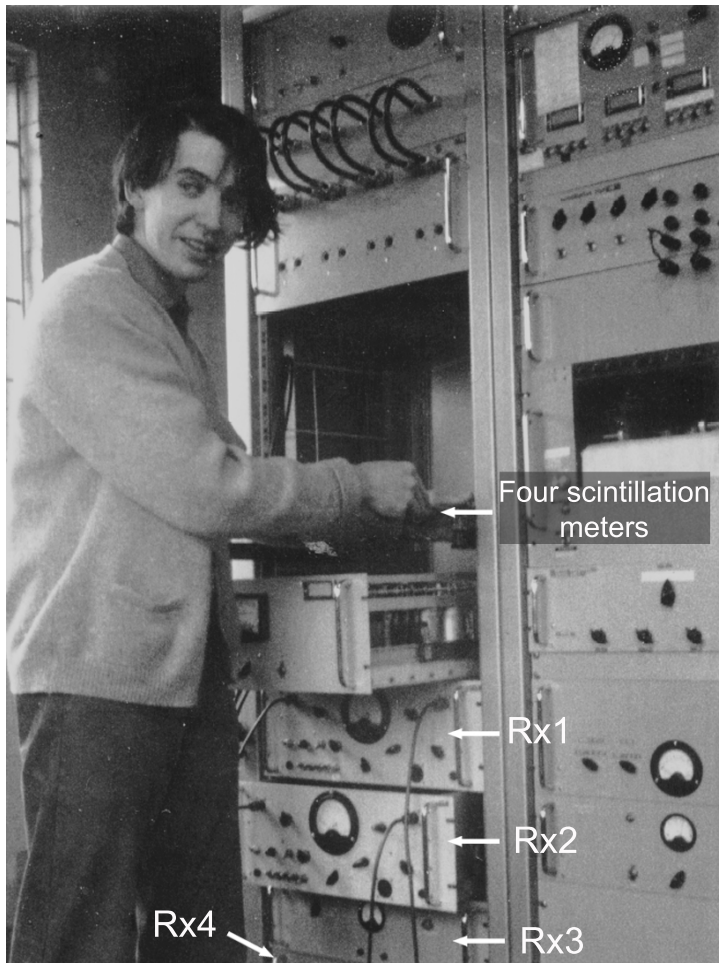

As can be seen in Fig. 2, the IPS signal from 4C 21.53 was very weak. The increased gain in the rectified and integrated high-pass filtered record was important in the detection of IPS from 4C 21.53. From the later IPS survey of Purvis et al., (1987), we know that the flux density of the pulsar in 4C 21.53 increased significantly between 1970 and 1979. So it was fortunate that it was bright enough in 1970 for IPS to be detected.

The next phase in the data processing was the addition of all the rectified and integrated high-pass filtered records. The scintillation survey (Readhead, 1972) ran continuously from 20 November, 1969, to 5 January 1971, with the whole sky between declination -8° and 80° being observed once per week, as it transited the 4-acre Telescope. Thus there were, in principle, 58 observations of the sky to be summed. In practice, ionospheric scintillation and radio frequency interference corrupted some of the observations, so that there were typically ~50 useable observations. The ionospheric scintillations, and almost all of the radio frequency interference, occurred at night, when the transiting objects were at large solar elongations, so the data lost to these effects was not as serious as it would have been during the day.

*Figure 3. The author installing the sliding rack shelf with the scintillation meters on 10 May, 1969*

I decided to digitize the data in order to combine the IPS observations from different days, for which purpose I built a separate channel for the four scintillation meters, using operational amplifiers, which simplified the design. Paul Scott very kindly installed analog to digital converters, and the integrated output was recorded on 7-track punch tape. It took 4 days to fill a 6" diameter paper tape, so by the end of the survey there were ~100 tapes. I had been writing the analysis pipeline and submitting frequent jobs to the University of Cambridge mainframe Titan computer, that served the whole university, and felt I was on top of the data analysis, but I was in for a nasty shock. The first time I submitted one of my large paper tapes for analysis; the tape was returned with a rip about 5 yards from the beginning of the tape and with ~4" of tape wrinkled and packed together. The tape had to be carefully smoothed out so as not to destroy any of the punched holes, and about 4 inches of black sticking tape, manufactured for this purpose, was pressed on to the paper tape to repair the tear. This was then placed, paper side up so that you could see the punch holes, over a precision die, and a second precision die that had dowels to align it with the lower die was then placed on top of the tape. A precision punch tool was then used to punch through all the original holes, thus salvaging the data. A second tape suffered the same fate. The problem, the computer operators explained, was that the computer was multiplexing between many jobs, so that it would read a few yards of tape, with the tape reader running at about 10 feet per second, and then stop as it switched to another job. To stop the tape rapidly a metal brake came down at high speed, and if there was a tiny piece of lint on the paper tape, due to a slightly imperfectly punched hole, the brake would catch the lint and the tape would be ripped. The unusually large paper tapes of the IPS survey meant that they would be stopped literally dozens of times while being read, and the chances of catching a stray piece of lint were very high. I discussed the problem with Judy Bailey, who helped to run the computing center and who also, conveniently, worked with the radio group. The least busy time on the computer, Bailey said, was between 4 a.m. and 5 a.m. So she suggested that I should come in and run my own jobs through the paper tape reader at that time. This required getting up at 3 a.m. and driving 12 miles to Cambridge, but within two weeks all of the paper tapes had successfully been read. While some of the tapes still did get ripped and require surgery, ~90% of them were read without any problems.

There were three stages in the analysis of the rectified integrated trace, which are described in detail in Readhead (1972). The results from the digitized, summed, integrated records, reproduced here from my thesis, for the two beams covering 4C 21.53 are shown in Fig. 4. In the two plots each of the six traces covers $4^h$ in Right Ascension. As can be seen here 4C 21.53 is clearly identified as a scintillating source in both beams, and the peaks are of comparable strength relative to the interpolated baseline on either side, showing that the source lies roughly half-way between the beam centers. To identify the scintillating sources with sources in the 4C catalog, I made a large map of the sky and compared the strengths of the signals with the positions and 178 MHz flux densities of the 4C sources. Since the scintillating sources were seen on one beam, or on two adjacent 4-acre Telescope beams, it was easy to estimate the declination and identify the 4C sources. The 4-acre Telescope was confusion limited at the lower end of the 4C catalog flux density scale, and any confusing 4C sources were noted. In the IPS survey 79 scintillating sources were found that were not identified with 4C counterparts, these unidentified scintillators were marked "U" in the catalog of Readhead and Hewish, (1974).

A key development, that occurred towards the end of the survey, was that I realized that the signal-to-noise ratio, when integrating or summing scintillation records, increases only as the fourth root of the integration time, and the fourth root of the number of records summed, (Readhead and Hewish, 1974), rather than as the square root, as had been assumed until then. Hewish and Ryle were both stunned when I informed them of this. This unexpected blow meant that the 4-acre Telescope did not have the sensitivity to observe IPS in all of the ~4,800 4C sources, as had been the design intent, but in only ~1500.

In setting out to search for my old chart records in April 2024, I contacted Malcolm Longair, who maintains close ties with the Cavendish Laboratory and the Lord's Bridge Observatory, to ask his advice. Longair contacted Clive Shaw at the observatory and in May very kindly took me out to Lord's Bridge and the three of us mounted a search for my charts. Most of the old chart records had been thrown on a bonfire some years ago during a big clean-up of the observatory, but we were astonished to find that two boxes of my chart records had been preserved. While this was only ~3% of the total chart records from my survey, it was for the critical 23° 23' beam, one of the two 4-acre Telescope beams that covered 4C 21.53. It is possible that John Baldwin, who supervised the preservation of the records at Lord's Bridge when most of them were put to flame, kept these records because of the importance of the IPS observations of 4C 21.53. Unfortunately, however, only one third of the charts could be found. The results found in these surviving chart records are summarized in Table 1. As can be seen there, the IPS nature of 4C 21.53 is clearly seen in only 1 of the surviving 17 records, the one from 9 February 1970 shown in Fig. 2, but it is likely that many of the records from early December 1969 to early March 1970 would have been of similar quality to this the sole surviving low-elongation chart.

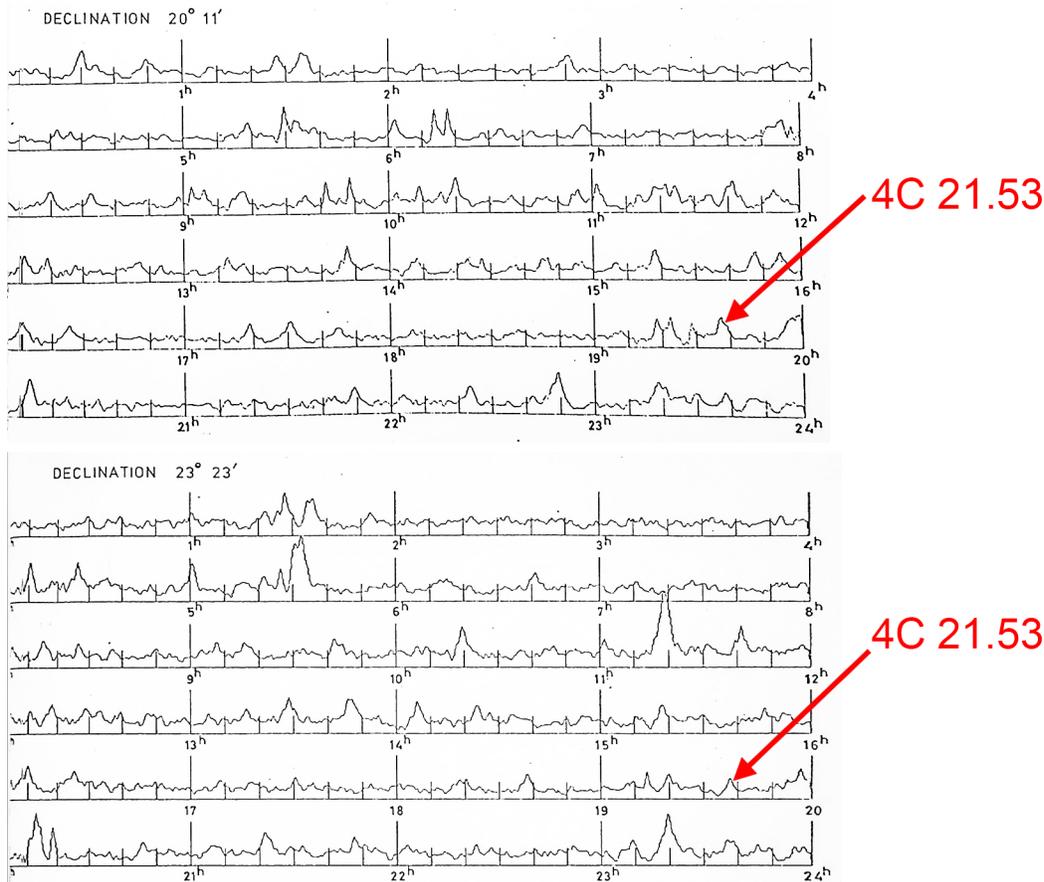

Figure 4. The summed integrated records showing 4C 21.53 in two beams. The ordinate is the scintillating flux. All records from a particular telescope beam of the bottom channel from Fig. 2 that were not corrupted by interference were added and then convolved with the integrated scintillating transiting beam pattern to produce these summed-convolved records. In this way the signal-to-noise ratio was increased by a factor 4, making weaker sources of IPS visible that were very difficult to see on the individual charts (Fig. 2).

| Date | Solar Elongation (degrees) | Comments |
| --- | --- | --- |
| 02/09/70 | 45.3 | clear signal in all three channels (see Fig. 2) |
| 04/06/70 | 78.1 | clean record, a hint of a signal in the integrated channel |
| 04/13/70 | 83.1 | clean record, a hint of a signal in the integrated channel |
| 04/20/70 | 88.2 | clean record, a small signal in the integrated channel |
| 04/22/70 | 89.6 | record reasonably clean, small peak in integrated channel |
| 04/27/70 | 93.2 | clean record, possibly a very faint hint of signal in integrated channel |
| 05/04/70 | 98.2 | noisy integrated channel for some reason, but there is a peak at the source position |
| 06/08/70 | 121.4 | clean record, a hint of a signal in the integrated channel |
| 06/15/70 | 125.4 | clean record, no clear scintillating signal |
| 06/22/70 | 128.9 | clean record, possibly a very faint hint of signal in integrated channel |
| 06/29/70 | 132.0 | clean record, a hint of a signal in the integrated channel |
| 07/06/70 | 134.5 | clean record, small peak in integrated channel |
| 07/13/70 | 136.4 | clean record, slightly higher average in integrated record |
| 07/20/70 | 137.4 | clean record, no clear scintillating signal |
| 07/27/70 | 137.7 | corrupted by radio frequency interference |
| 08/17/70 | 133.3 | clean record, a small signal in the integrated channel |
| 08/31/70 | 127.2 | corrupted by ionospheric scintillation |

Table 1. Summary of the surviving charts records of 4C 21.53.

When it came to analyzing my data in 1971, it appeared to me that there was a dearth of scintillating sources at low Galactic latitudes. This had to be approached with care because the background noise due to the Galactic synchrotron emission increases dramatically at low Galactic latitudes, thereby reducing the sensitivity of the telescope, and this had to be taken into account. I attributed the dearth of sources exhibiting IPS at low Galactic latitudes only partly to this loss of sensitivity, and the rest I thought was due to interstellar scattering.  In December 1971, Ryle suggested to me that I should set aside my thesis temporarily in order to write a paper on the interstellar scattering result.  I was trying to finish my thesis as fast as possible because my funding had run out, and my wife and I were living off her salary.  Ryle offered to support me for 6 months if I could persuade my college (St. Johns) to split the cost, which they generously agreed to. Accordingly, I received checks for £157 from both St. Johns College and the Cavendish Laboratory, totaling £314, which, in 1971, was sufficient to support a research student for 6 months.  For the next two months I worked on writing the paper based on my partially completed survey.  The paper was submitted to Nature in February 1972 (Readhead and Hewish, 1972).  This paper did not include 4C 21.53.  The reason was that I had not yet completed the analysis of my thesis data, and I was dubious about the scintillating signal from 4C 21.53, which was very weak.  I had found that the well-known radio source, 3C 274 (M87), had slipped through my analysis pipeline and appeared as a scintillating source. I had examined many chart records of 3C 274, and I knew that there was not a scintillating component in this object. I felt very strongly that it was important not to classify any dubious cases as definite scintillators, thereby running the risk of sending people off on wild goose chases for compact structure that did not exist, and for this reason I was deliberately conservative.  Years later Hewish asked me why I had been so conservative, and told me that all of the objects classified as probable scintillators in Readhead and Hewish (1974), had been found to be definite scintillators by Purvis et al., (1987). It is ironic, but easily understandable,  that the paper by Readhead and Hewish, (1972), which does not include 4C 21.53, is cited by Backer, et al. (1982), whereas the paper by Readhead and Hewish (1974), which does include 4C 21.53, is not cited by them. This was not a misperception, but simply an error.

I submitted my thesis in August 1972 (Readhead, 1972), but I was not satisfied that the results were publishable at that time because of the misidentification of some non-scintillating sources, such as 3C 274, as scintillators.  I therefore decided to spend the first year of my post-doctoral position, as the Royal Society Weir Research Fellow, reviewing the 5 miles of 3-track chart recordings and confirming all of the scintillators and non-scintillators by eye. It was an intense year, with chart records often spread all over the living room and the bedroom floors in the small converted stable where my wife and I were living in the village of Willingham 12 miles out of Cambridge. During this year one of the sources I confirmed as a definite scintillator, based on the combined integrated output (Fig.4) and the chart records (Fig. 2), was 4C 21.53. It is important to note that, had 4C 21.53 not been a 4C source, I would still have identified this as a scintillating source, the only difference being that it would have been one of the  "unidentified" scintillating sources listed in Readhead and Hewish (1974).

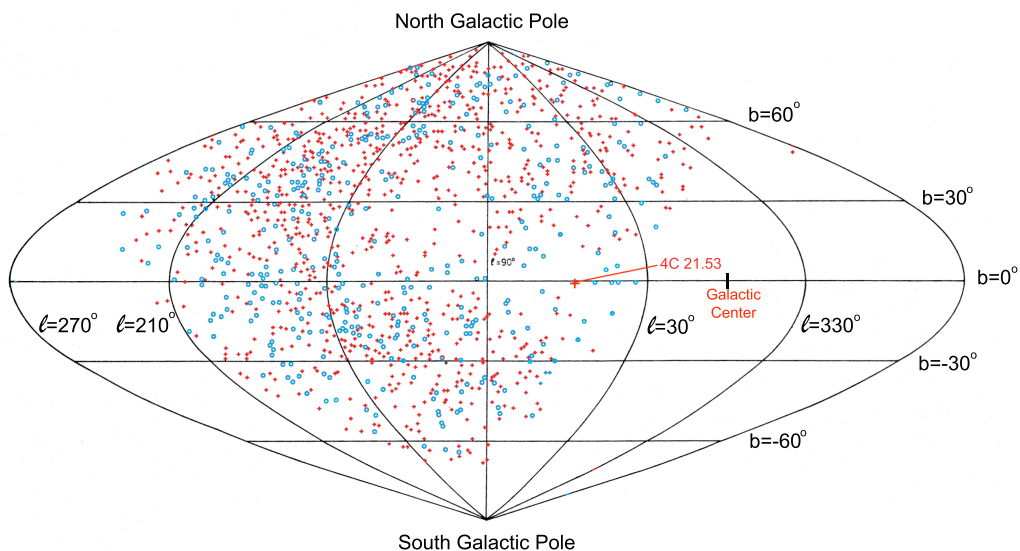

Figure 5. The distribution of the 1500 sources at 81·5 MHz: $R \geq 0·4$ (+) and $R < 0·4$ (o).

The next phase of the IPS story at Cambridge was that I re-analyzed the numbers of scintillating sources using my completed survey Readhead and Hewish (1974), and, together with Peter Duffett-Smith, carried out an IPS study comparing the angular sizes at 81.5 GHz, from my thesis, with observations at 151 MHz carried out by Duffett-Smith of 32 strong scintillating sources, that had low minimum solar elongations, which did not include 4C 21.53 (Duffett-Smith and Readhead, 1976). The 151 MHz observations were carried out using the fixed section of the 4C Telescope (Pilkington and Scott, 1965), and, apart from helping as an extra pair of hands, when needed with the observations, I focused on the re-analysis of the statistics from my thesis, while Duffett-Smith, who was an excellent instrumentalist, focused on the 4C Telescope 151 MHz observations. The distribution of the strongly-scintillating, and the weakly- and non-scintillating, sources is shown in Fig. 5, which is adapted from Duffett-Smith and Readhead, (1976). The strongly scintillating sources are indicated by red crosses, and 4C 21.53 is highlighted by a large red cross. The weakly-scintillating and non-scintillating sources are marked with blue circles. The anomalous nature of 4C 21.53 is immediately apparent in this plot, since this is the only strongly scintillating (R>0.4) source at very low Galactic latitude (b = -0.3°) and Galactic longitude ~60°, i.e. it is the only strong scintillator in this region of sky within a few degrees of the Galactic plane.

It was this aspect of the IPS in 4C 21.53 that drew so much attention to this object, as described in Demorest et al. (2024). In my re-analysis of the statistics of the scintillating sources in my survey (Duffett-Smith and Readhead, 1976), I found that the effects of interstellar scattering (see Appendix) where considerably weaker than reported in Readhead and Hewish (1972), but I still found a small effect. In a re-analysis of the Cambridge IPS data, Rickard and Cronin (1979) found no statistically significant evidence for increased interstellar angular broadening in the Galactic plane, and they suggested that 4C 21.53 belonged to a possible new class of radio source, which they named "scintars". The discussion of the path after 1976 towards the second crucial step in the discovery of millisecond pulsars is covered in detail in the companion paper by Demorest et al. (2024).

**CONCLUSION**

The scintillating flux density reported by Readhead and Hewish (1974) was 4.2 Jy at elongation 45°. In Purvis et al., (1987) it is 7.5 Jy at this elongation. Thus the flux density of the millisecond pulsar 1937+214 increased by a factor 1.8 between the two surveys. This level of flux density variation, which is thought to be caused by refractive interstellar scintillation, is typical in PSR 1937+214 (Ramachandran et al., 2006). In Fig. 6 we show the flux density of PSR 1937+214 at 327 MHz observed by these authors from 1995 to 2005, during which time it changed by a factor four between 1999 and 2004. Had the same fractional variations pertained at 81.5 MHz between 1970 and 1979, then in 1970 the scintillating flux density could well have been below 2 Jy, and hence below the level of detectability on the 4-acre Telescope.

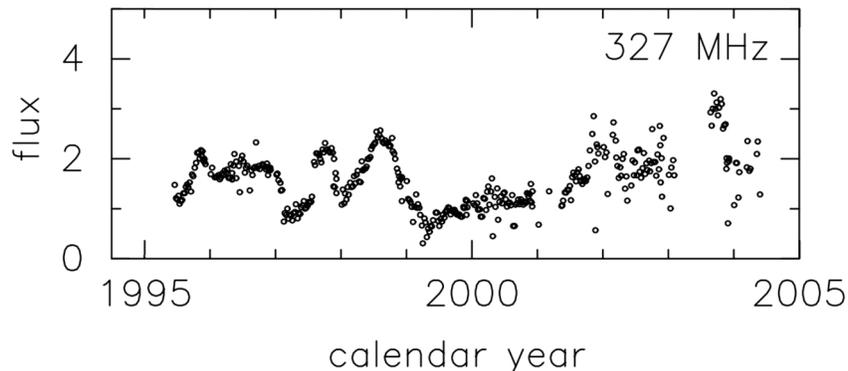

*Figure 6. The 327 MHz flux density of PSR 1937+214, adapted from Ramachandran et al., (2006).*

It was fortunate that the pulsar was bright enough to be detectable in my survey, but the signal-to-noise ratio on individual days was low. It is for this reason that only an upper limit on the angular size (<0.95 arc seconds) was measured and that it was classified as a "Class C" scintillator – the lowest category of definite scintillators.

In conclusion, it is easy for misperceptions to arise, even for people totally immersed in a field, and, in the excellent accompanying article by Demorest et al. (2024), there is a highly significant example:

"At the 1984 Green Bank Millisecond Pulsar Workshop, Backer pointed out this amazing coincidence regarding 4C21.53: "The 4C instrument would have bypassed 1937+214 had it not been for the coincidental arrangement of three unrelated objects [radio sources] of comparable strength in two of the 4C interferometer lobes. Any one of the sources would not have been above the 4C intensity threshold. If 1937+214 had been isolated on the sky, we might not be here today!"

This is an ironic misperception because the fortuitous random juxtaposition of sources in the 4C catalog actually obscured the chain of events that led to the discovery of millisecond pulsars. The Cambridge IPS survey of Readhead and Hewish (1974), was deliberately designed as a "blind" survey, and was in no way based on the 4C catalog, although the original goal of the IPS survey, was to search for IPS in all of the objects in the 4C catalog, but this was not possible because the sensitivity of the 4-acre Telescope was much lower than had been thought due to the fourth root factor I discovered that was described earlier. Being blind, the IPS survey was unbiased as to whether or not the scintillating sources detected were 4C sources. The identification with 4C sources was carried out only after the blind scintillation survey had been completed, and there was no expectation that all of the scintillating sources would be identifiable with 4C sources. The IPS from 1937+214 would therefore have been discovered and catalogued in Readhead and Hewish (1974), had there been no identification with a 4C source, along with the other 79 unidentified sources marked as "U" in their catalog. Its position, halfway between the 20º 11' and 23º 23' beams, and more accurate position in right ascension, due to the dimensions of the 4-acre Telescope, placed it at RA 19h 37m, and Dec 21º 47' in epoch 1950 coordinates, which may be compared with the position of 4C 21.53W of RA 19h 37m 29.6s, and Dec 21º 30' 33.0" measured at the VLA by Rickard et al, (1983). Thus the position observed in the 4-acre survey revealed the very low Galactic latitude, and therefore the anomalous nature of this object. This would have drawn attention to it as a very unusual unidentified compact source, even if the 4C survey had not included it. There can be no doubt that observations would have followed to determine an accurate position, so we can rest assured that Don Backer and colleagues would have been at the Green Bank Millisecond Pulsar Workshop in 1984, even had there not been a coincidental arrangement of confusing sources in the 4C catalog.

## ACKNOWLEDGEMENTS

I thank Jocelyn Bell-Burnell, Roger Blandford, Malcolm Longair, and Sterl Phinney, for encouraging me to write this historical account, and Malcolm Longair and Clive Shaw for their assistance in searching for the IPS chart records from my IPS survey, and for recovering sufficient of such as to make Fig. 2 and Table 1 possible. I thank Jocelyn Bell-Burnell, Malcolm Longair, Tim Pearson, and Sterl Phinney for many comments and suggestions that have greatly improved the text and the figures of this article. I thank Allen Packwood and Jessica Saunders, of the Churchill College Archives, for their help in locating the 4-acre Telescope log books, and providing copies of the portions covering the period of my thesis. I also thank Shri Kulkarni, Miller Goss, and Paul Demorest for making contacts that were essential for this article, for their enthusiasm for a complete account of the discovery of millisecond pulsars, and for permitting me to read their article prior to publication.

## APPENDIX: INTERPLANETARY AND INTERSTELLAR SCINTILLATION AND SCATTERING

Interplanetary and interstellar scintillation refers to variations in the intensity of electromagnetic waves, that are caused by variations in the density of the electrons in a plasma through which the electromagnetic wave is traveling. The electron density variations cause changes in the refractive index, and hence the speed, of electromagnetic waves transiting the plasma. This leads to interference between electromagnetic waves travelling along different paths through the plasma, which causes scintillation. In interplanetary scintillation, the inhomogeneities are in the solar wind. In interstellar scintillation they are in the more tenuous interstellar medium.  Just as planets do not twinkle because they are larger than the angular scale of the bending of light from stars, so only sufficiently compact radio sources show interplanetary scintillation (IPS), and therefore in the days before radio interferometry with continental baselines, IPS was used to distinguish very compact radio sources (like pulsars) from larger ones (like the giant radio lobs in galaxies with black holes) https://en.wikipedia.org/wiki/Interplanetary_scintillation.

There are two types of scintillation: "diffractive" and "refractive", that are explained in the review by Rickett (1990). The interplanetary scintillation used by the author in his thesis is weak and is due to diffractive scintillation, whereas the scintillation that causes the brightness of PSR 1937+214 to vary by large amounts is refractive scintillation.

Scattering refers to the angular broadening of a source caused by the same electron density variations that cause strong scintillation, and it occurs in both the interplanetary medium and the interstellar medium. Significant interplanetary scattering occurs only for objects whose line of sight to earth passes very close to the sun, which is not relevant for this article. However interstellar scattering caused by electron density variations in the interstellar medium can broaden the apparent size of radio sources at low galactic latitudes, where the path length through the Galactic plasma is high, to the point where their apparent size is too large to cause IPS.